# Topological carnival: electrically-powered motions of toron crystallites in chiral liquid crystals

Hayley R. O. Sohn[1] and Ivan I. Smalyukh[1,2,3*]

[1]Department of Physics and Materials Science and Engineering Program, University of Colorado, Boulder, CO 80309, USA

[2]Department of Electrical, Computer and Energy Engineering and Soft Materials Research Center, University of Colorado, Boulder, CO 80309, USA

[3]Renewable and Sustainable Energy Institute, National Renewable Energy Laboratory and University of Colorado, Boulder, CO 80309, USA

*Email: ivan.smalyukh@colorado.edu

*Malleability of metals is an example of how dynamics of defects like dislocations induced by external stresses alters material properties and enables technological applications. However, these defects move merely to comply with the mechanical forces applied on macroscopic scales whereas the molecular and atomic building blocks behave like rigid particles. Here we demonstrate how motions of crystallites and defects between them can arise within the soft matter medium in an oscillating electric field applied to a chiral liquid crystal with polycrystalline quasi-hexagonal arrangements of self-assembled topological solitons called "torons". Periodic oscillations of electric field applied perpendicular to the plane of hexagonal lattices prompt repetitive shear-like deformations of the solitons, which synchronize the electrically-powered self-shearing directions. The temporal evolution of deformations upon turning voltage on and off is not invariant upon reversal of time, prompting lateral translations of the crystallites of torons within quasi-hexagonal periodically deformed lattices. We probe how these motions depend on voltage and frequency of oscillating field applied in an experimental geometry resembling that of liquid crystal displays. We study the inter-relations between synchronized deformations of the soft solitonic particles and their arrays and the ensuing dynamics and giant number fluctuations mediated by motions of crystallites, 5-7 defects pairs and grain boundaries in the orderly organizations of solitons. We discuss how our findings may lead to technological and fundamental science applications of dynamic self-assemblies of topologically protected but highly deformable particle-like solitons.*

**Significance**: *Topological solitons exhibit particle-like behavior and, similar to colloids, allow one to model out-of-equilibrium processes like crystallization, melting and defect motions in molecular and atomic systems. We now show that such soft solitonic particles can also exhibit dynamics not accessible to their atomic and molecular counterparts. These dynamics arise from facile responses of the liquid crystal host medium to external fields, causing non-reciprocal shearing-like deformations of quasi-hexagonal lattices of torons in a periodically tilted director background. Collective motions of crystallites of these solitons prompt fascinating evolution of grain boundaries and 5-7 defects within them. Our findings reveal rich dynamic behavior of topologically protected objects at both the building block and crystal lattice scales.*

Soft matter systems often exhibit behaviors intermediate between that of crystals and fluids, with a broad range of emergent phenomena arising from a plethora of competing interactions that are typically weak and comparable in strength to thermal fluctuations [1,2]. The building blocks of these systems are often soft in a sense that their size can change dramatically with tuning temperature, like in the case of specially designed polymer particles [3], or particle dimensions can be even tuned by electric fields, like in the case of topological solitons in liquid crystals (LCs) [4,5]. Such colloidal and solitonic soft matter systems have been widely used to model dynamic behavior in their atomic and molecular counterparts, including crystallization and melting [3-7], dynamics of defects within crystalline lattices [7-9], formation of glasses and gels [6,8-10], and so on. An open remaining question is how and under what conditions the soft nature of the building blocks of soft matter can lead to emergent dynamic behavior that is inaccessible to solid-state systems. Here we describe such unexpected behavior in crystalline lattices of topological solitons dubbed "torons" [11,12] in chiral LCs under conditions that resemble those in LC displays [1].

LC torons are energetically-stable spatially-localized structures in the ordering direction of LC's rod-like molecules, which is described by the so-called "director" field **n(r)** [11,12]. Topologically similar structures were also found in non-centrosymmetric solid-state magnets [13] and in their magnetic colloidal counterparts [14,15]. The elementary torons studied here can be understood as the elementary two-dimensional $\pi_2(\mathbb{S}^2/\mathbb{Z}_2) = \mathbb{Z}$ skyrmions (low-dimensional analogs of Skyrme solitons in nuclear physics [12, 16]) terminating at $\pi_2(\mathbb{S}^2/\mathbb{Z}_2) = \mathbb{Z}$ point defects (LC analogs of the magnetic Bloch points [15,16]) near the confining substrates [17]. Similar to the case of magnetic systems [18,19], the point defects that are elements of the same homotopy group serve as a means of confining these two-dimensional (2D) solitonic structures within a finite three-dimensional (3D) volume, making skyrmion fragments of finite length [19], which in our case is defined by the gap between solid glass substrates confining the LC material. In this work, we study how torons in chiral nematic LC morph in response to an oscillating external electric field **E**, as well as how they move within periodically self-shearing crystalline lattices while prompting a complex dynamic evolution of grain boundaries separating them. A combination of polarizing video microscopy, optical imaging and numerical modeling through minimization of free energy reveals 3D structures of **n(r)** and provides insights into the physical origins of the observed phenomena. These findings uncover the richness of behavior of topological solitons and their responses to periodic external stimuli accompanied by motions of periodically-shearing assemblies of crystallites. The similarity of realization conditions and voltage driving with those used in LC displays [20] may lead to new technological applications relying on emergent electro-optic behavior.

**Results**

At high packing densities, hexagonal arrays of torons resemble the A-phase of skyrmions in chiral magnetic systems (Fig. 1 and SI Fig. S1) [21-23], although here the skyrmion within each toron terminates on point defects near surfaces of a thin LC film (Fig. 2) and can be observed with an optical microscope (Fig. 1B-D). No fields are needed for such crystallites of torons to exist, but a voltage $U$~1 V periodically modulated electric field applied orthogonally to the sample plane (Fig. 1A,B) leads to complex emergent behavior that quickly culminates in translational motions of the crystallites along roughly the same spontaneously chosen direction (Fig. 1C,D and SI Video S1). When this motion develops, the torons move forward and backward in anti-parallel directions upon turning the square-wave voltage on and off within each period, though the magnitudes of the frequency-dependent back-and-forth translations within each period $T$ are different and result in a net motion (Fig. 1D,E). By sweeping the modulation frequency within $f=1/T$=1-1000 Hz (Fig. 1E,F), we note that the coherent motions of torons can be reversed, similar to the behavior of individual solitons [17]. Associated with morphing of **n(r)** within each toron in a complex non-reciprocal way, the magnitudes of the lateral translations in the anti-parallel directions depend on $f$, so that the relative forward (or backwards) shifts of toron positions are larger (or smaller) at frequencies <100 Hz (or >100 Hz), leading to the reversal of motion directions at $f$~100 Hz while sweeping frequency and keeping other parameters unchanged (Fig. 1E). This behavior is consistent with the characteristic 50-100 ms response times of LCs to electric fields in this experimental geometry [17]. With motions starting at $U$>1.5 V (Fig. 1G), the average velocity of toron motions increases with the voltage amplitude $U$, though more complex structural transformations take place at $U$>2.5 V, which tend to destroy periodic lattices of torons and are beyond the scope of our present study [24].

Computer-simulated structures of individual torons reveal the $\pi_2(\mathbb{S}^2)$ ($\pi_2(\mathbb{S}^2/\mathbb{Z}_2)$) topology of the vectorized (nonpolar) **n(r)**-field, both in terms of the skyrmion tube orthogonal to the plane of the LC sample and the point defects on which it terminates near substrates (Fig. 2). These computer simulations also allow one to define preimages (regions of constant orientation of the LC director) corresponding to the north and south poles of the $\mathbb{S}^2$ order parameter space of vectorized **n(r)** and effectively defining torons as particle-like objects (Fig. 3). The elementary torons have the skyrmion number ±1 of the 2D topological soliton, which matches the hedgehog charge of the ±1 point defects at confining surfaces on which the skyrmion tube terminates to match the topologically trivial perpendicular boundary conditions. The signs of these topological invariants depend on the direction of vectorization of the LC's nonpolar **n(r)** and switches to opposite upon the reversal of vectorization direction [12]. Preimages of all points cross the toron's midplane and also terminate on both point defects that serve as the sources/sinks of the vectorized field lines (Fig. 2A,D,K). When such torons self-organize into crystalline lattices, they remain as spatially localized topological particles both at no fields (Fig. 2B,E,G,H, L) and when external electric field is applied (Fig. 2C,F,I,J,M). Experimental and computer-simulated polarizing optical micrographs (Fig. 2G-J) show that voltage application morphs the roughly axi-symmetric **n(r)** structure within each toron (Fig. 2B,E), making it asymmetric in both in-plane cross-sections parallel to the background far-field director **n**$_0$ perpendicular to substrates (Fig. 2C,F) and vertical cross-sections containing **n**$_0$ (Fig. 2M). At $U$ =1.5-2.5 V, the point defects near opposite confining surfaces shift asymmetrically and the spontaneously chosen direction of tilting of **n(r)** correlates between different torons within the lattice (Fig. 2C,F,M). While this complex behavior of 3D toron structures within crystals is difficult to capture by simply visualizing **n(r)** within cross-sectional planes (Fig. 2), it can be followed and correlated between multiple torons within crystallites by probing behavior of the singular point defects and preimages of vectorized **n(r)** (Fig. 3). At no fields, the preimages and point defects of torons within crystals look like particles in orderly hexagonal assemblies (Fig. 3B,C,G,L,M) of individual toron counterparts (Fig. 3A,F,K). However, the singular points shift and preimages morph with applying $U$ (Fig. 3D,E,I,N,O). This behavior is very consistent with the voltage-induced evolution of polarizing optical micrographs obtained between parallel polarizers (Fig. 3H,J), where this polarizing microscopy setting is selected to visualize geometry and location of the effective preimages of north and south poles of the order parameter space as bright regions. Remarkably, both modeling and experiments reveal that the torons within lattices are effectively sheared (Fig. 3L-O). This electrically-powered self-shearing of torons is accompanied by tilting of south-pole preimages and corresponding lateral shifts of the singular point defects in opposite directions in a plane orthogonal to the motion direction (Fig. 3L-O).

Since both the Reynolds and Ericksen numbers are low for our system [1,2], translational motion of crystallites of torons requires that the evolution of **n(r)** and/or flows within the LC medium are not invariant upon reversal of time within the effective on and off "strokes" of each $T$. To get insight into how this happens, we have probed the temporal evolution of textures using polarizing optical video microscopy (Fig. 4A and SI Video S2). The north-pole preimages, effectively defining torons as quasi-particles, adopt shapes of deformed and partially inter-merged hexagons which rotate synchronously with the voltage modulation (Figs. 3 and 4). The period of this preimage rotation is consistent with *T*, though there is a slight lagging in response to instantaneous voltage changes caused by relatively slow response of the complex 3D **n**(**r**) to periodically modulated *U*. Importantly, as voltage is effectively turned on and off within each *T*, the toron's preimages rotate in different directions, counterclockwise and clockwise, respectively (Fig. 4). The magnitudes of angles of these opposite rotations are different too, so that the director evolution that is manifested through such textural evolution is not invariant upon reversal of time (Fig. 4B-F). Consequently, each toron within the lattice also translates, with opposite positional shifts upon turning voltage on and off but with a net translation along the motion direction as a result of voltage modulation within each *T*.

Probing the details of crystallite motions reveals fascinating behavior of torons and the defects within them as they exhibit complex coherent dynamics (Fig. 5). The toron crystallites translate while preserving their quasi-hexagonal order. Torons undergo thermal fluctuations within the lattice while exhibiting a net translation in the same spontaneously chosen direction (Fig. 5A-D), with the locally meandering but long-term straight trajectories of toron motions being statistically non-distinguishable (Fig. 5B,C). The net long-term linear displacement of torons, averaged over the field of view, is proportional to time when modulated electric field is applied, but averages to zero at no fields (Fig. 5E). At the same time, the mean square displacement (MSD) is linear in time at no fields and scales quadratically with time at applied fields that power motions of toron crystallites (Fig. 5F), with the emergent motion taking place along a well-defined spontaneously-selected direction for all torons within a crystallite (Fig. 5A-F). Interestingly, the synchronized motions of torons are accompanied not only by periodic non-reciprocal modulations of toron field configurations, but also by correlated motions of singular point defects within the torons (Figs. 3O and 5G), which can be tracked with unpolarized transmission-mode video microscopy when the microscope's focal plane is adjusted to coincide with the sample's plane containing point singularities (insets of Fig. 5E and SI Fig. S2). The point defects in this imaging mode are visible as small dots because of the light scattering on them due to the reduced scalar order parameter and the corresponding localized variations of the LC medium's effective refractive index. While the thermally driven lateral step displacements of point defects at no fields are described by Gaussian-like distributions and appear to be laterally aligned with the thermal fluctuations in positions of torons overall, this behavior changes dramatically upon applying modulated *U* (Fig. 5G). The lateral distribution of the tracked point defect positions within the toron become shifted with respect to the moving geometric center of the electrically morphed (with a period of *T*) toron, with the distributions in the plane containing the velocity vector **v** $\parallel$ **x** being broader and with smaller displacement with respect to toron's geometric center as compared to such a distribution in the plane orthogonal to **v** (Fig. 5G). The skyrmion tubes within torons appear to be (on average) somewhat compressed in the direction of motion and stretched in a direction orthogonal to it, with applied voltage and motion effectively deforming the originally hexagonal lattice of "soft" reconfigurable-particle-like torons (Fig. 5H). As the applied voltage increases, the distribution of the toron stretching directions becomes more and more narrow, peaking at 90° relative to **v** (Fig. 5H). This anisotropic out-of-equilibrium electrostriction is accompanied by morphing of toron lattices within $T$ and motions of grain boundaries and 5-7 disclination defects within the crystallites on larger time scales $>T$.

The complex motions of the squishy toron particles and defects within their lattices give origins to giant number fluctuations (Fig. 6A). Using video microscopy, we analyze the mean $<N>$ and root mean square $\Delta N = <(N-<N>)^2>^{1/2}$ of torons within different square-shaped sample areas containing defects within toron lattices. Torons within the crystallites with grain boundaries exhibit giant number fluctuations with $\Delta N \propto <N>^\alpha$, where $\alpha=0.762$ is obtained from fitting (Fig. 6A). This is consistent with fluctuations in the local number density of torons probed by counting the instantaneous numbers of torons within a selected square-shaped sample area versus time (Fig. 6B), as well as with motions of lattice defects in the deformed skyrmion lattices visualized using the Voronoi construction (Fig. 6C-E, SI Fig. S3 and SI Video S3). While the directions of the motions of 5-7 defect pairs, both standalone and within percolating grain boundaries (Fig. 6C-E and Fig. 7), do not exhibit strong correlations with respect to **v**, the grain boundaries tend to shrink and the crystallites anneal with time as motion progresses (Fig. 7A-F), making the positional correlations in the radial distribution function more and more long-ranged (inset of Fig. 7B, SI Fig. S4).

To quantitatively examine how motion affects the orientational ordering, we have characterized the local bond orientational order parameter defined as $\psi_j = \frac{1}{n_j}\sum_{k=1}^{n_j} e^{im\theta_{jk}}$ (see Methods), where $m = 6$ for hexatic order. While $\psi_j$ often approaches unity, the hexatic order parameter decreases on a larger scale, $\Psi_6 = \left|\frac{1}{N}\sum_{j=1}^{N}\psi_j\right|$ (see Methods), due to the misalignment of the quasi-hexagonal domains of torons (Fig. 7G), though it tends to increase even within larger areas as motion progresses (Fig. 7G-I). Within the same areas (Fig. 7J-L and SI Fig. S5), the motion directions of individual torons are poorly correlated at the onset (Fig. 7G and SI Fig. S5A) but become synchronized with time (Fig. 7H,I and SI Fig. S5B,C). Most of the regions of the sample's field of view exhibit significantly deformed lattices of torons, with deformations in area larger than 2% depicted in colors coding the direction of stretching relative to **v** (Fig. 7J-L). Clearly, most of the crystallites of torons are stretched in the direction roughly orthogonal to **v** (Fig. 5H and Fig. 7J-L) while being compressed along **v**. This out-of-equilibrium dynamics is, however, rather complex because the motions of grain boundaries and other defects within lattices also influence the electrically induced periodic self-shearing behavior. Moreover, a sample's imperfections like dust particles and torons anchored on such imperfections also influence this behavior, so that the deformations of toron lattices are more complex within certain sample areas. The velocity order parameter, defined as $S = |\sum_j^N \mathbf{v}_j|/(N\, v_c)$, characterizes the degree of ordering of velocity vectors $\mathbf{v}_j$, where $N$ is the number of skyrmionic particles and $v_c$ is the absolute value of velocity of a coherently-moving crystallite. $S$ increases from the onset of motion throughout the first couple minutes of motion (SI Fig. S5), where it then approaches the system's dynamic equilibrium at $S \approx 0.65$, significantly lower than the $S$ values of schools of sparse skyrmions without positional correlations during motion [24]. On the other hand, the fact that motion of torons becomes coherent within short period of time and leads to increased ordering correlations, evidenced by the characterization of hexatic order parameter, is intriguing and consistent with the behavior of more fluid-like schools of skyrmions studied previously [24].

**Discussion and conclusions**

Typically, crystalline solids are thought to be incompatible with motions of building blocks on large scales (generally associated with fluid condensed matter systems), though malleability of solids is an example of how large-scale dynamics can be mediated by motions of dislocation defects. Recent interest in active matter brought about concepts of "frozen flocks" and active "solids" [25-29], which may relate to familiar behaviors of directed crowds and traffic jams. However, the out-of-equilibrium dynamic behavior of crystallites in material systems are poorly understood and rarely studied. Our findings show that soft-particle-like topological solitons can exhibit electrically powered motions that stem from energy conversion on individual particle level but then translate to large-scale coherent motions of crystallites of torons accompanied by periodic self-shearing-like deformations. Unlike solid hard-sphere-type colloids, atoms or most other building blocks of condensed matter, these solitons display electric reconfigurations and morphing due to transformations within the lattices and grain boundaries between the crystallites. While recently studied active systems were paralleled with every-day life examples like polarized crowds and traffic jams [26], the elaborate toron motions with 2D meandering of skyrmion tubes and 3D coherent configurational dynamics of both skyrmion tubes and point defects within torons packed into crystals reminds of the moving-while-dancing crowds in carnivals.

Although the toron lattices remain quasi-hexagonal in nature, with the local hexatic order parameter being rather high, the application of field and motion make individual torons behave like weakly asymmetric polar particles, which synchronize their orientations and motions with time (Fig. 7). Such reconfigurations of symmetry are not easy to achieve within more common types of building blocks of matter, like colloids, especially within the crystalline lattices that they form. However, qualitatively similar effects could potentially arise due to ordering of Janus-like active particles [30]. Moreover, crystallites could potentially even form in very dense crowds of people or herds of animals, which could be deformable or prone to shearing, though not as squishy as torons. Therefore, it may be of interest to explore how our "crowded" assemblies of dynamic inanimate solitonic particles would compare to the out-of-equilibrium behavior of other physical systems, including those of biological origin. Up to date, however, such motions of crystals of active particles have not been achieved beyond what we describe here, at least to the best of our knowledge.

To conclude, we have demonstrated that topological solitons exhibit electrically powered emergent dynamics so far not accessible to their colloidal, atomic and molecular counterparts. These dynamics arise from facile responses of the liquid crystal host medium to external fields, and particularly from the response to turning voltage on and off that is not invariant upon reversal of time. Such complex responses of this soft matter system effectively convert injections of energy by electric pulses at the individual toron level into motion, leading to emergent motions of entire crystallites. Our experiments revealed how collective motions of crystallites of these solitons prompt fascinating evolution of grain boundaries and 5-7 defects within them. Since the studied behavior emerges in samples and under conditions resembling those in LC displays, we anticipate that our findings may lead to new applications in electro-optics, photonics, displays and diffractive optics. The rich emergent behavior of singular point defects and the solitonic skyrmion tubes within torons self-assembled into crystallite arrays can be paralleled with such elaborate and complex crowd motions of dancing-while-moving carnivals.

**Materials and Methods**

**Sample preparation and generation/manipulation of torons**

To realize torons experimentally, a chiral nematic ZLI2806 (EM Chemicals) was doped with right-handed chiral dopant CB-15 (EM Chemicals) at a weight fraction $C_{\text{dopant}} = 1/(\xi \cdot p)$ to define the helicoidal pitch $p$ of the chiral LC, where $\xi$ is the helical twisting power of the chiral dopant (Table S1) [5]. The chiral nematic was additionally mixed with 0.1wt% of cationic surfactant Hexadecyltrimethylammonium bromide (CTAB, purchased from Sigma-Aldrich) to allow for spontaneous generation of torons by means of relaxation from electrohydrodynamic instability [24]. The samples were prepared by sandwiching these mixtures between indium tin oxide coated glass substrates. Strong perpendicular boundary conditions were set for the LC director field by treating the substrate glass with polyimide SE1211 (Nissan Chemical) by spin coating it at 2700 rpm for 30s, followed by a 5 min pre-bake at 90°C and a 1 h bake at 180°C. Samples of thickness $d$=10μm were assembled using glass spacers. In addition, commercial cells (purchased from Instec) were used, with the same thickness and boundary conditions but also with patterned indium tin oxide electrodes defining the area within which torons are initially generated. The solitons were generated by first inducing and then relaxing electrohydrodynamic instability (SI Fig. S1) obtained at $U$=20 V and $f$=2 Hz, forming spontaneously as energetically favorable structures after turning $U$ off because of the chiral LC's tendency to twist [24]. By manually switching on and off $U$ that induces the hydrodynamic instability 3-5 times in a matter of a few seconds, one can increase the number density as desired, up to tight packing of torons. The initial locations of the as-generated torons are random, but crystallites slowly form due to repulsive interactions at high packing densities (SI Fig. S1). Electric field was applied across the samples using a homemade MATLAB-based voltage-driving program coupled with a data-acquisition board (NIDAQ-6363, National Instruments) [17], which was done in order to morph the solitons and power crystallite motions via macroscopically-supplied energy.

**Numerical modeling**

The Frank-Oseen free energy functional describes the energetic cost of spatial deformations of **n(r)** within a chiral nematic LC:

$$F = \int d^3\mathbf{r} \left\{ \frac{K_{11}}{2} (\nabla \cdot \mathbf{n})^2 + \frac{K_{22}}{2} [\mathbf{n} \cdot (\nabla \times \mathbf{n})]^2 + \frac{K_{33}}{2} [\mathbf{n} \times (\nabla \times \mathbf{n})]^2 + K_{22} q_0 \mathbf{n} \cdot (\nabla \times \mathbf{n}) \right.$$
$$\left. - K_{24} \{\nabla \cdot [\mathbf{n}(\nabla \cdot \mathbf{n}) + \mathbf{n} \times (\nabla \times \mathbf{n})]\} \right\}, \quad (1)$$

where the Frank elastic constants $K_{11}$, $K_{22}$, $K_{33}$, and $K_{24}$ describe the energetic costs of splay, twist, bend, and saddle-splay deformations, respectively and $q_0 = 2\pi/p$ characterizes the LC chirality. Strong boundary conditions (consistent with experiments) on the surfaces are assumed, excluding surface energy from our calculation. We assumed $K_{24}$=0 or $K_{24}$=$K_{22}$ [11,12], in both cases obtaining qualitatively similar results, and all other elastic constants utilized in numerical modeling are based on experiments (Table S1). When an external electric field is applied, Eq. (1) is supplemented with the corresponding electric field coupling term:

$$F_{\text{electric}} = -\frac{1}{2} \int d^3\mathbf{r} (\mathbf{E} \cdot \mathbf{D}) = -\frac{1}{2} \int d^3\mathbf{r} (\mathbf{E} \cdot \bar{\bar{\varepsilon}} \mathbf{E}), \quad (2)$$

where **E** is the applied electric field, **D** is the electric displacement field in the dielectric LC medium and $\bar{\bar{\varepsilon}}$ is the dielectric tensor with components $\varepsilon_{ij} = \varepsilon_0(\varepsilon_\perp \delta_{ij} + \Delta\varepsilon n_i n_j)$, where $\varepsilon_0$ is the vacuum permittivity, $\varepsilon_\perp$ is the perpendicular dielectric constant measured when electric field is applied perpendicular to the director, and $\Delta\varepsilon$ is the dielectric constant anisotropy (Table S1).

Under our experimental conditions, the toron field configurations emerge as local or global minima of the total bulk free energy given by the sum of Eqs. (1) and (2), yielding the numerically generated structures at various applied fields. A variational-method-based relaxation routine is used to perform numerical modeling of the energy-minimizing **n(r)** [4,11,12]. At each iteration of the numerical simulation, **n(r)** is updated based on an update formula derived from the Lagrange equation of the system, $n_i^{new} = n_i^{old} - \frac{MSTS}{2}[F]_{n_i}$, where the subscript $i$ denotes spatial coordinates, $[F]_{n_i}$ denotes the functional derivative of F with respect to $n_i$, and MSTS is the maximum stable time step (representing the time between iterations) in the minimization routine, determined by the values of elastic constants and the spacing of the computational grid [11]. The stopping condition is found by monitoring the change in the spatially averaged functional derivatives, which, as it approaches zero, indicates that the system is in a state corresponding to the energy minimum and the relaxation routine is terminated. The 3D spatial discretization is performed on large 3D square-periodic 128×128×32 grids, and the spatial derivatives are calculated using finite difference methods with second-order accuracies. This allows us to minimize discretization-related artifacts in modeling of the structures of these topological solitons. To construct a preimage of a point on $\mathbb{S}^2$ within the 3D volume of a topological soliton, we calculate a scalar field defined as the difference between the solitonic field **n(r)** and a unit vector defined by the target point on $\mathbb{S}^2$. The preimage is then visualized with the help of the isosurfaces of a small value in this ensuing scalar field [4,12]. The effective physical dimensions of torons as particle-like objects were defined as sample regions that are interior of north-pole preimages corresponding to torons (Fig. 3), whereas the effective toron's "particle" surface was defined as the surface separating the north-pole far-field preimage from other preimages. Computer-simulated polarizing optical micrographs (Fig. 2H,J) were generated using a Jones matrix method [31,32] in MATLAB (obtained from MathWorks), in which the configuration of optical axis ≡**n(r)** is sampled layer by layer through the experimental cell thickness, pitch, and optical anisotropy (Table S1).

**Optical microscopy, video characterization and data analysis**

All experimental images and videos were captured using polarizing or bright-field transmission-mode optical microscopy using an Olympus BX-51 upright microscope equipped with charge-coupled device cameras either Grasshopper (Point Grey Research, Inc.) or SPOT 14.2 Color Mosaic (Diagnostic Instruments, Inc.) and dry 4x, 10x, and 40x objectives (numerical apertures ranging from 0.3 to 0.9). In bright-field optical micrographs, the point defects are visible as distinct sharp dot-like features because they strongly scatter light when microscope's focal plane coincides with the plane containing these defects. The videos were analyzed for positions of torons and point defects using open-source software ImageJ/FIJI's (National Institute of Health) particle tracking capabilities coupled with a software plugin "wrMTrck" for generating positional data for each toron, extracted frame by frame with motion. The position and particle-counting data were exported to MATLAB to characterize trajectory pathways and displacements, giant-number fluctuation scaling, density fluctuations, and various order parameters. Toron number density (extracted using the particle-counting tools in ImageJ) was used to characterize giant-number fluctuations with time for thirty areas of different sizes, ranging from 15μm x 15μm to 1500μm x 1500μm, for each experimental video. A log-log plot of the mean $<N>$ and root mean square $\Delta N=<(N-<N>)^2>^{1/2}$, shown in Fig. 6A, was obtained using the compiled density data points by area. One such representative number density fluctuation trend that was used for a single point in Fig. 6A is shown in Fig. 6B for a 2000μm$^2$ region. A composite of 5 videos were analyzed for different regions within the samples, resulting in ~150 data points from which the scaling trend was extracted. Fluctuations were characterized for each video over time periods of 500-600 s.

The positional data for each topological soliton was coupled with the MATLAB function *knnsearch* to define nearest neighbors, compare positions, create Voronoi diagrams using the *voronoin* function, and analyze shearing behavior, packing, and bond orientational order. The local hexatic bond orientational order $\psi_j$ [33-35] was used to color-code each individual toron in Fig. 7G-I. This analysis was then expanded to calculating the hexatic order parameters for different sample areas (Fig. 7G-I). The velocity order parameter $S$ was calculated using the evolution of the positional data between the consecutive frames, where the velocity vector $\mathbf{v}_j$ was defined by drawing a vector, pointing in the direction of motion, between each toron's position in consecutive frames of the experimental video. [36] These velocity vectors were then used to determine $S = |\sum_j^N \mathbf{v}_j|/(N v_c)$ for $N$ particles within the field of view and average magnitude of crystallite velocity, $v_c$ (SI Fig. S5).

**ACKNOWLEDGEMENTS.** We acknowledge discussions with P. Ackerman, C. Marchetti, B. Senyuk, J.-S.B. Tai and M. Tasinkevych. This research was supported by the National Science Foundation through grants DMR-1810513 (research), DGE-1144083 (Graduate Research Fellowship to H.R.O.S.) and ACI-1532235 and ACI-1532236 (RMACC Summit supercomputer used for the numerical modeling).

# Figures and Legends

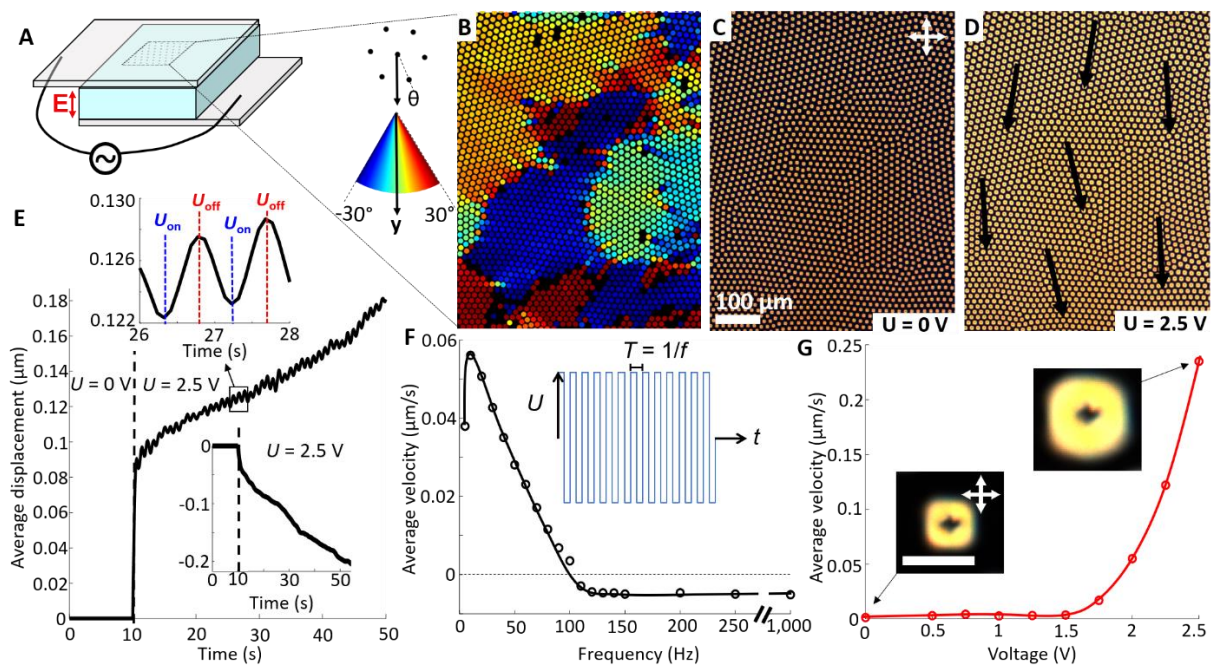

**Fig. 1.** Generation and dynamics of crystals of torons. (*A*) Schematic of a sample with voltage application across the LC by using transparent electrodes on the inner surfaces of the confining substrates. The direction of the periodically oscillating alternating electric field **E** is marked in red. (*B-D*) Close-packed crystallites of torons are shown colored according to orientations (*B*) and in polarizing micrographs at no applied voltage (*C*) and at $U$=2.5 V (*D*), where the black arrows denote motion directions of each crystallite. The color scheme for visualizing crystallite orientations is shown in the inset between (*A*) and (*B*). (*E*) Average displacement of torons in each crystallite with time, where $U$=2.5 V with carrier frequency of 1 kHz is turned on and then square-wave modulated at 1 Hz. The top inset shows the details of back-and-forth displacements that emerge with voltage switching at instances marked with $U_{on}$ and $U_{off}$, corresponding to turning voltage on and off, respectively. The bottom inset shows similar displacement at the 1 kHz carrier signal with no modulation (note the much slower motion in an opposite direction). (*F*) Crystallite velocity dependence on the modulation frequency of **E**, with a schematic of the square-wave voltage profile given in the inset. (*G*) Velocity dependence of crystallite motions on voltage, with insets showing polarizing optical images of torons at $U$=0 and $U$=2.5 V. The scale bar is 10 μm. Crossed polarizer orientations are marked with white double arrows throughout.

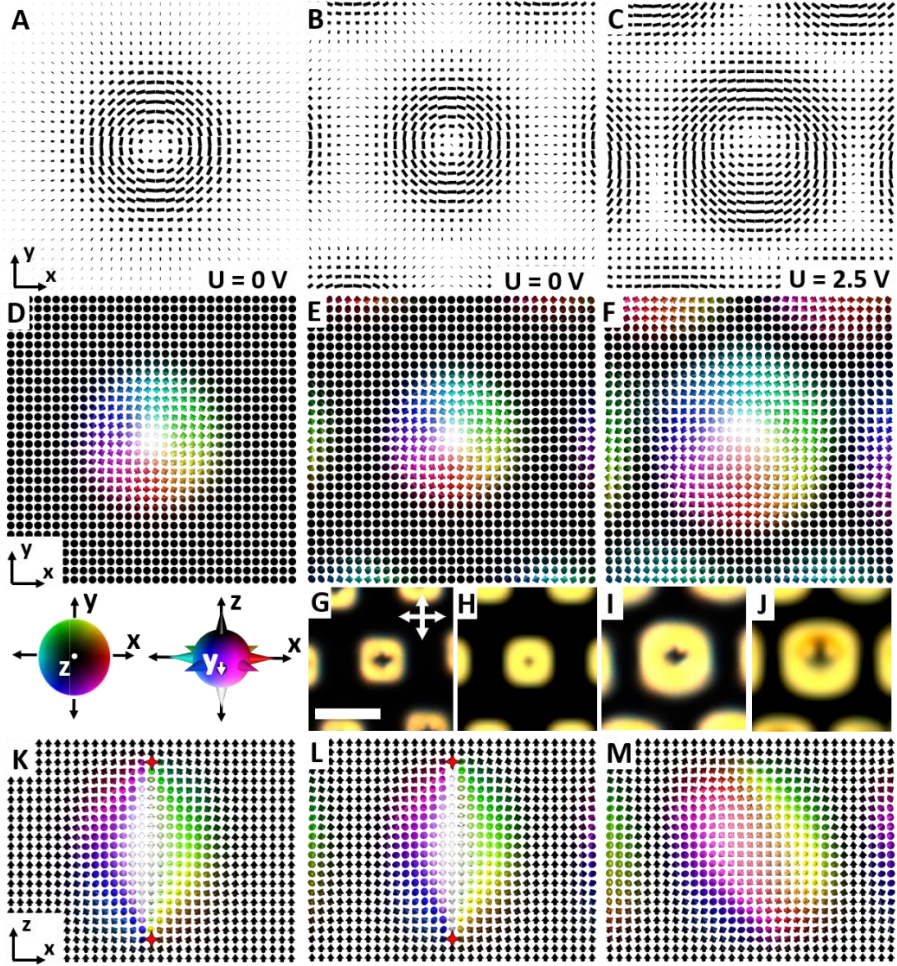

**Fig. 2.** Topology and electric reconfiguration of torons. The left column represents an individual toron at $U$=0 V, the middle column represents a toron within a crystallite at $U$=0 V, and the right column represents a toron within a crystallite at $U$=2.5 V. (*A-F*) Numerically-simulated **n(r)** of an individual toron (*A* and *D*) and torons within quasi-hexagonal crystallites at $U$=0 V (*B* and *E*) and $U$=2.5 V (*C* and *F*) shown as a director structure in the x-y midplane as black rods (*A-C*) and (*D-F*) as vectorized **n(r)** colored according to points on $\mathbb{S}^2$ shown in the bottom inset of (*D*). (*G-J*) Polarizing optical images of torons within a quasi-hexagonal crystallite, obtained experimentally (*G* and *I*) and simulated numerically (*H* and *J*), respectively. White double arrows denote the crossed polarizer orientations and the scale bar is 10 μm. (*K-M*) Cross-sectional x-z midplanes of vectorized **n(r)** colored according to the $\mathbb{S}^2$ scheme in (*D*) for a single toron (*K*) and torons within a quasi-hexagonal crystallite at no applied field (*L*) and at applied field (*M*). Red crosses in (*K* and *L*) denote the hyperbolic point defects; similar point defects are not visible in (*M*) because voltage-powered shearing of the toron pushed them out of this vertical cross-section.

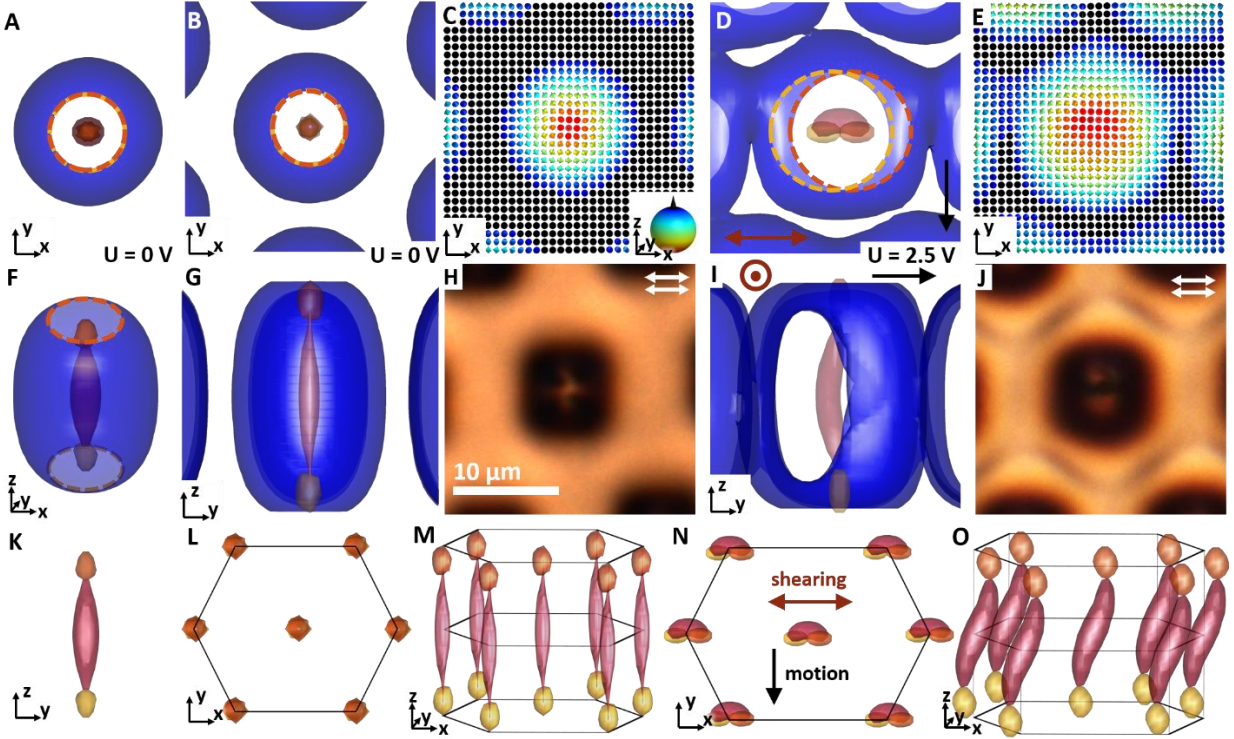

**Fig. 3.** Dynamics of toron lattices. The first left-side column represents an individual toron at $U=0$ V, the next two columns represent a toron within a hexagonal crystallite at $U=0$ V, and the last two columns represent a toron within a hexagonal crystallite at $U=2.5$ V. (*A-E*) Computer simulations showing blue surfaces separating particle-like marching torons from the uniform far-field director $\mathbf{n}_0$ exterior (roughly corresponding to north-pole preimages), south-pole preimages (magenta) shown in the x-y view (*A,B,D*) and vectorized-$\mathbf{n}(\mathbf{r})$ within the x-y midplanes (*C,E*). Arrows are colored according to their orientations and $\mathbb{S}^2$ in the inset of (*C*) for vectorized $\mathbf{n}(\mathbf{r})$. Orange and yellow dashed circles mark the edges of the surface effectively confining the toron's interior and separating it from the background with alignment along $\mathbf{n}_0$. (*F-J*) The same surfaces shown in the vertical y-z views (*F,G,I*) and experimental parallel-polarizer transmission-mode optical images (*H* and *J*). Polarizer orientations are marked with white double arrows. (*K-O*) South-pole preimages (magenta) and point defects (orange and yellow, corresponding to top and bottom singular point defects, respectively) of an individual toron (*K*) and a hexagonal unit cell of torons shown from the x-y view (*L* and *N*) and the 3D perspective view (*M* and *O*). Directions of stretching/shearing and motion upon voltage application are marked with the maroon and black arrows, respectively (*D*, *I*, and *N*). Axes of the common coordinate system are shown throughout.

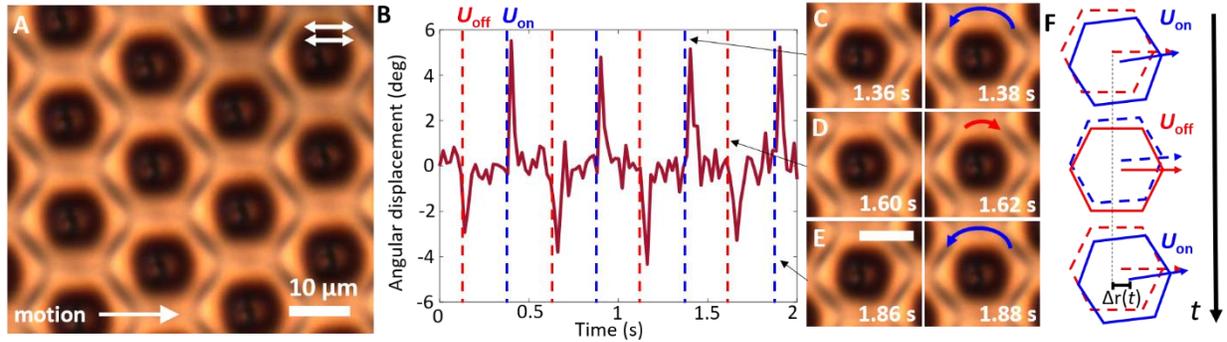

**Fig. 4.** Nonreciprocal structural transformations of crystals of torons during motion. (*A*) Parallel-polarizer experimental optical image of a toron lattice during motion at $U$=2.5 V and with carrier frequency 1 kHz modulated at 2 Hz. Polarizer orientations are marked with white double arrows and the direction of motion is shown in the bottom-left corner. (*B-E*) Nonreciprocal angular dynamics of a single toron within a crystallite upon voltage modulation (*B*), with corresponding experimental images (*C-E*) showing the toron immediately before (left-side image) and after (right-side image) each instantaneous voltage change. Curved blue and red arrows denote corresponding directions of rotation. The scale bar is 10 μm. (*F*) Schematics of angular rotations at different times corresponding to motion, with the ensuing net lateral displacement labeled as Δr(t). Red and blue symbols are used throughout to represent the temporal instances when the instantaneous voltage is turned off ($U_{off}$) and on ($U_{on}$), respectively.

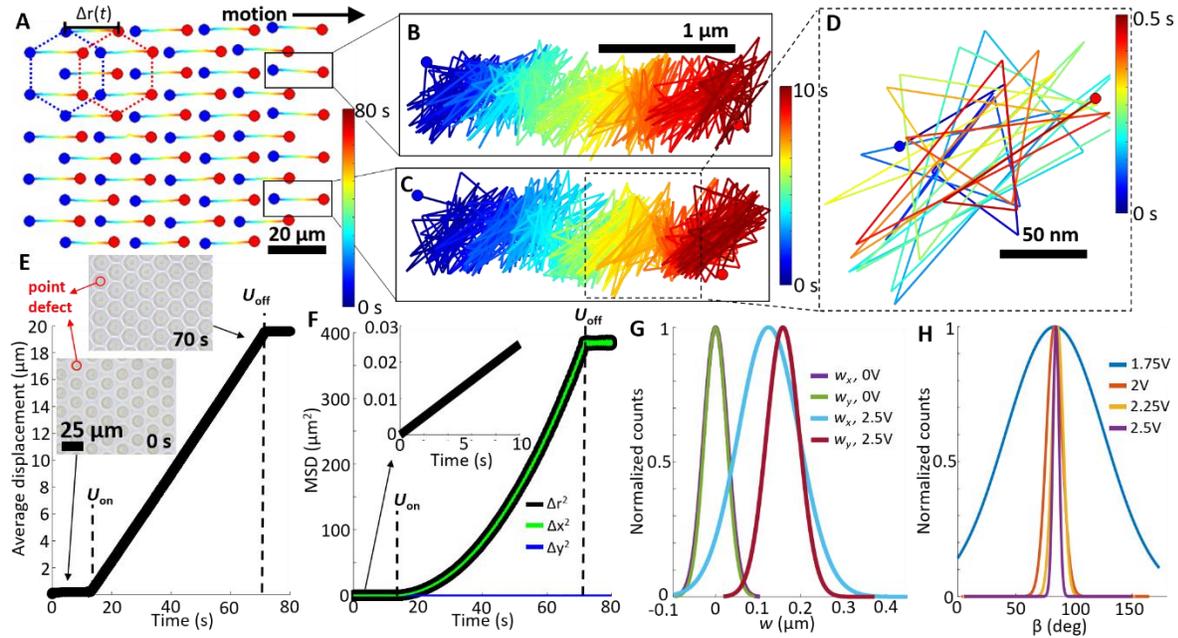

**Fig. 5.** Dynamics of torons and the singular point defects within them. (*A-D*) Trajectories of toron motions within a crystallite at $U$=2.5 V and $f$=10 Hz, progressively zooming in on the details of motion, and shown colored according to elapsed time (with the maximum elapsed time marked in each part). Direction of motion, net displacement $\Delta r(t)$, and colored hexagons indicating the hexagonal unit cell before and after motion are marked in (*A*). (*E*) Average displacement of the hyperbolic point defects near the upper confining substrate, analyzed with bright field microscopy (see the video frames in the insets, with red circles encircling a point defect in each frame). (*F*) Mean squared displacement (MSD), plotted both as net displacement ($\Delta r^2$) and displacement in two orthogonal directions ($\Delta x^2$ and $\Delta y^2$). The inset shows a zoom-in on $\Delta r^2$ in the first ten seconds, before voltage has been switched on. Dotted lines marked with $U_{on}$ and $U_{off}$ mark where $U$=2.5 V with $f$=10 Hz modulation is switched on and off, respectively (*E* and *F*). (*G*) Experimental normalized distributions for the distance from the geometric toron center to upper point defect ($w$) measured along x- and y-directions parallel and perpendicular to **v**, respectively, sampled over ~0.02s between the frames. (*H*) Experimental normalized distributions for the angle, β, between the direction of toron lattice stretching and the net direction of motion at different voltages.

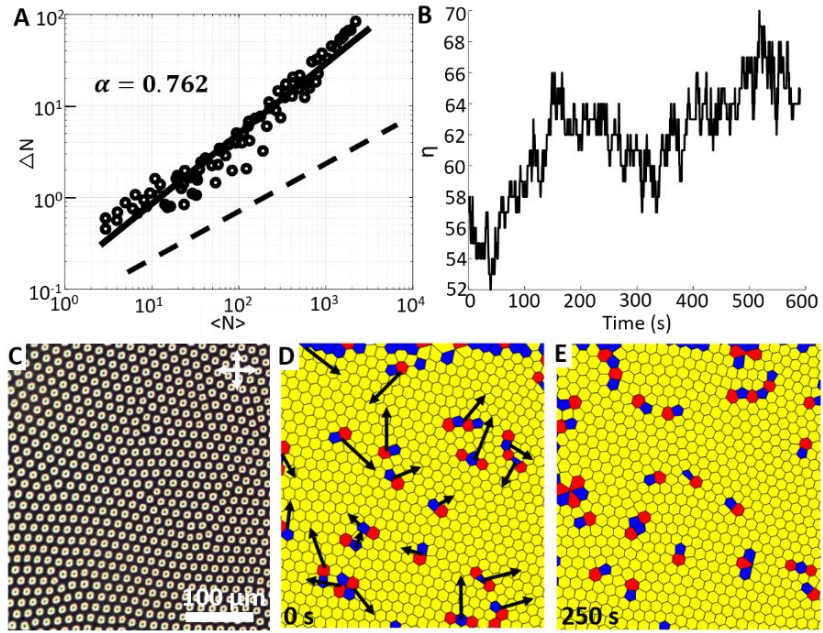

**Fig. 6.** Number density and giant number fluctuations in squishy toron lattices. (*A*) A log-log plot of ΔN versus <N>; black dashed line indicates a slope of 0.5 for reference. (*B*) An example of number density fluctuation revealed by counting the number of torons η during motion while passing a 2000μm² sample area. (*C-E*) Polarizing image of toron crystallites (*C*) with corresponding reconstructed Voronoi diagrams at times 0s (*D*) and after 250s of motion (*E*). The Voronoi diagrams are colored according to each toron's number of nearest neighbors (5 = blue, 6 = yellow, 7 = red). Black arrows in (*D*) indicate the directions of motion of lattice defects. Crossed polarizer orientations are marked with white double arrows in (*C*).

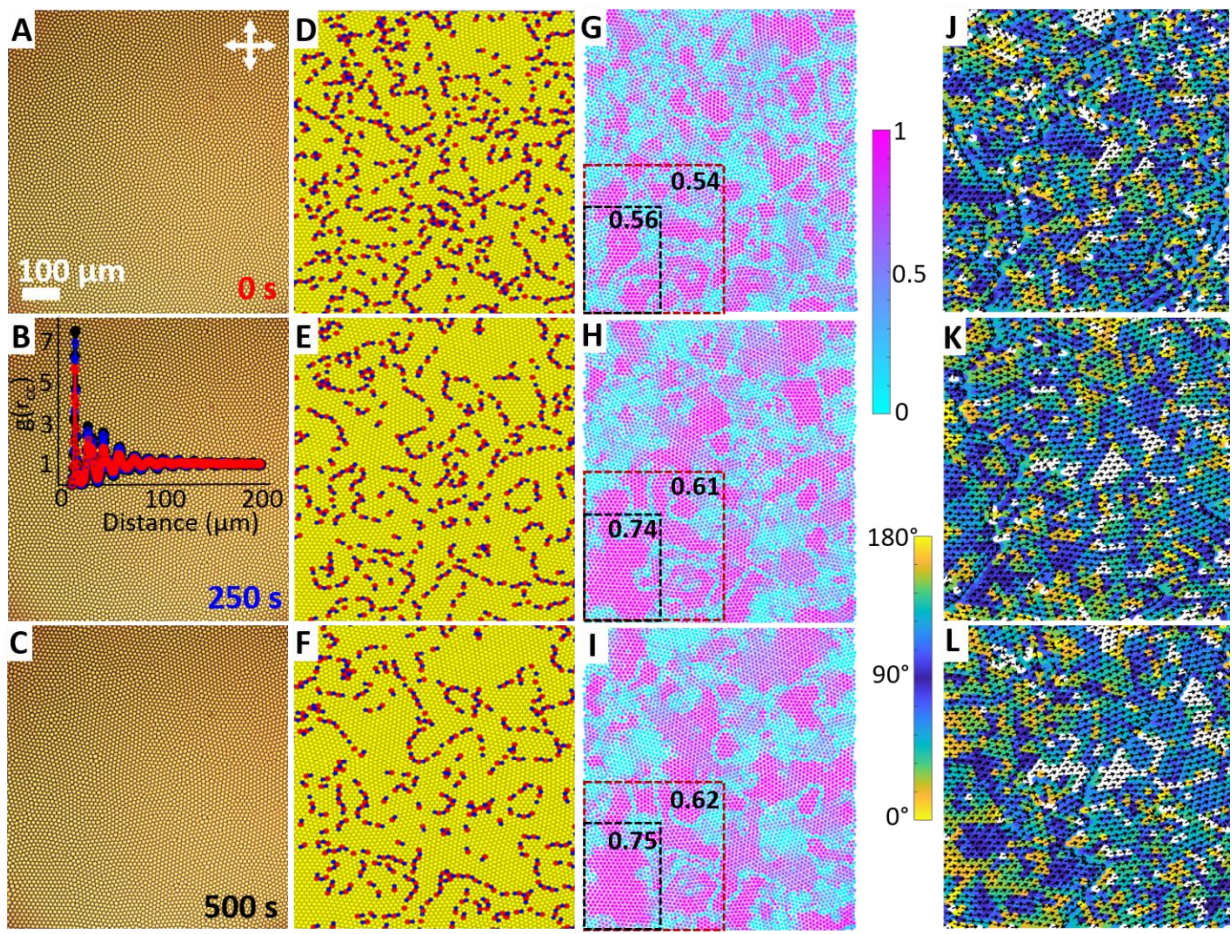

**Fig. 7.** Evolution of grain boundaries, bond orientational order, and velocity order with motion. (*A-C*) Polarizing images of a large field view with toron crystallites during motion at $U=2$ V and $f=50$ Hz, with the inset in (*B*) showing the radial distribution functions g($r_{cc}$) for each image, colored according to the color of the timestamps on corresponding micrographs. Crossed polarizer orientations are marked with white double arrows. (*D-F*) Voronoi diagrams constructed using the experimental images (*A-C*) and colored according to the number of nearest neighbors (5 = blue, 6 = yellow, 7 = red). (*G-I*) Experimental images for each toron in the moving crystallites colored according to degree of local hexatic bond orientational order parameter according to the scheme in the right-side inset of (*G,H*). Dashed-line boxes are marked with the hexatic order parameter values for the corresponding areas. (*J-L*) For the areas within each image outlined by the maroon dashed boxes in (*G-I*), directions of motion are marked with black arrows for each toron, with the background color coded according to the angle $\beta$ (color scheme presented in the inset on the left side of *L* and *K*), representing the relationship between individual unit cell stretching and the average left-to-right direction of motion. The areas that are left white correspond to quasi-hexagonal crystallite regions with lateral stretching <2%.

# Supporting Information

**Table S1. Material properties of the nematic host and chiral additive**

| Material/Property | $\Delta\varepsilon$ | $K_{11}$ (pN) | $K_{22}$ (pN) | $K_{33}$ (pN) | $\Delta n$ | $\xi$ of CB15 ($\mu m^{-1}$) |
|---|---|---|---|---|---|---|
| ZLI2806 | -4.8 | 14.9 | 7.9 | 15.4 | 0.044 | +5.9 |

Experimental dielectric anisotropy, $\Delta\varepsilon$, elastic $K_{11}$, $K_{22}$, $K_{33}$ constants, optical anisotropy, $\Delta n$, of the nematic host mixture and helical twisting power, $\xi$, for a right-handed chiral additive CB15 utilized in numerical modeling.

# Supplementary Figures and Legends

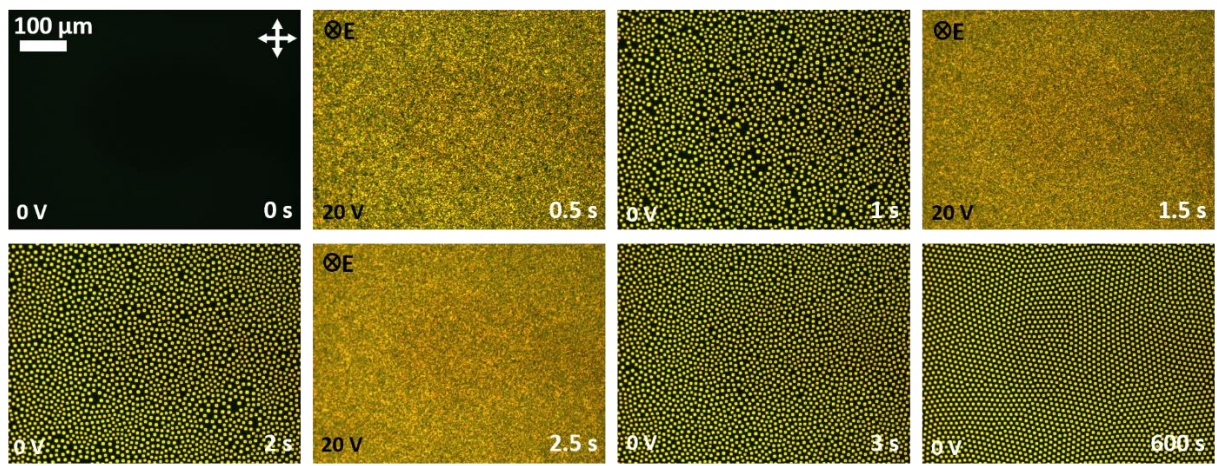

**Fig. S1.** Generation of dense crystallites. Polarizing optical images show the generation of large number densities of tightly-packed torons via the relaxation from electrohydrodynamic instability at $U$=20 V and $f$=2 Hz following repetitive turning voltage on and off. Relaxation of the sample over time at no applied fields then leads to large hexagonal crystallites of tightly-packed torons. The instantaneous voltages in the image series are marked in the bottom-left parts of the images. The elapsed time is marked in the bottom-right parts of the images. **E** is orthogonal to image and sample planes.

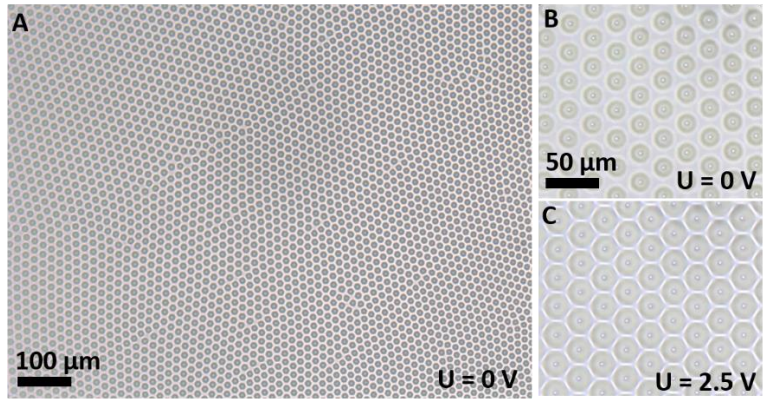

**Fig. S2.** Point defect visualization and tracking. (*A-C*) Bright-field transmission-mode optical micrographs of the upper hyperbolic point defects at $U=0$ (*A,B*) and $U=2.5$ V (*C*). The images are obtained by co-locating the microscope's focal plane with the sample's plane containing point defects.

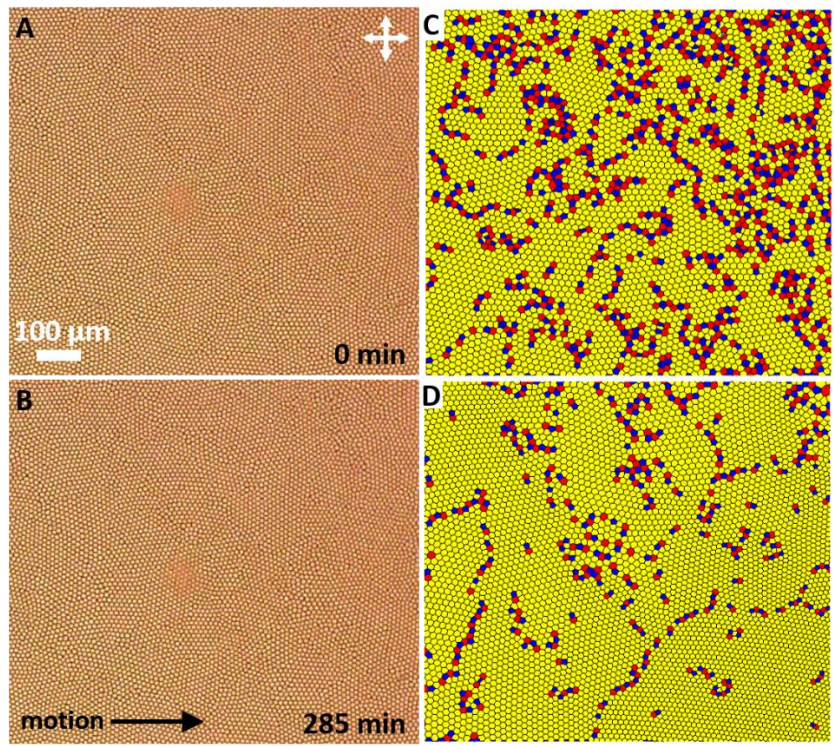

**Fig. S3.** Crystallite evolution with motion over a few hours. (*A* and *B*). Polarizing optical images of a large field-of-view of the LC sample with toron crystallites during motion at $U=2$ V and $f=50$ Hz. Crossed polarizer orientations are marked with white double arrows. (*C* and *D*) Voronoi diagrams corresponding to the images shown in (*A* and *B*), colored according to the number of nearest neighbors (5 = blue, 6 = yellow, 7 = red).

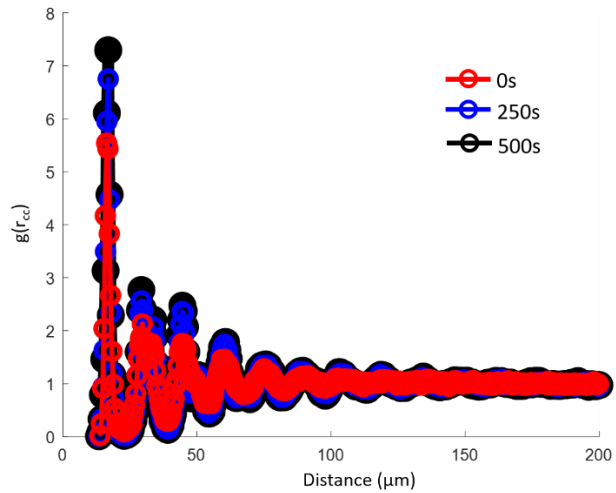

**Fig. S4.** Radial distribution functions for crystallites at different elapsed times, corresponding to the Fig. 7B inset.

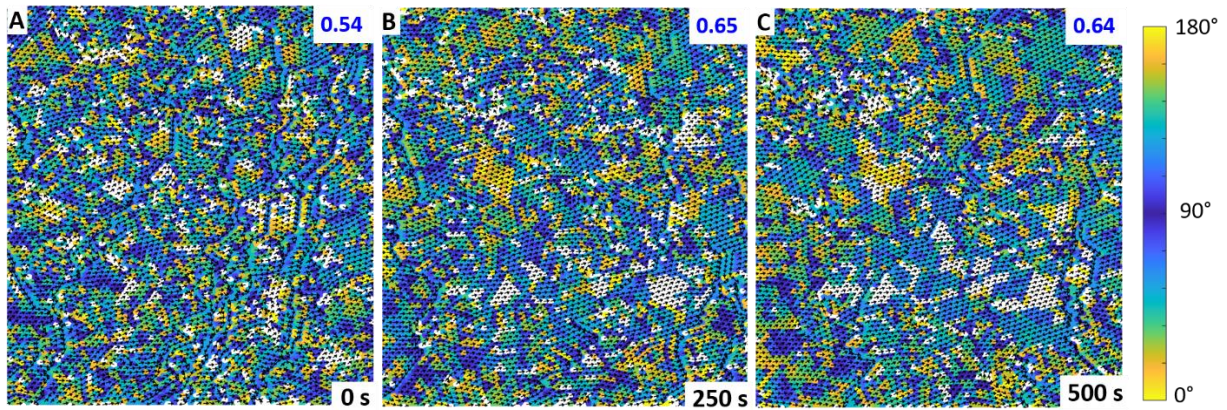

**Fig. S5.** Velocity vector fields and β for sample's large fields of view for moving toron crystallites. (*A-C*) Evolution of velocity vector fields corresponding to the images shown in Fig. 7A-C. Background color represents the angle β (right-side inset of *C*). The corresponding values of the velocity order parameter *S* are marked in the top-right of the figure parts. The elapsed time is shown in the bottom-right of the same figure parts. The areas that are left white correspond to the crystallite regions with lateral stretching <2%.

# Supplementary Video Captions

**Video S1:** Polarizing optical video of crystallite motion at $U$=2.5 V and $f$=10 Hz. The average direction of motion is marked with a white arrow and the crossed white double arrows denote polarizer orientations. The video is played at ~60x speed and the total elapsed time is ~ 7 minutes.

**Video S2:** Polarizing optical video of nonreciprocal response to voltage modulation that results in crystallite motion when voltage is switched on at $U$=2.5 V and $f$=2 Hz. Parallel polarizer orientations are marked with white double arrows, direction of motion is labeled in the bottom left, and the on/off states of voltage application are marked for each frame in the upper right corner. The video is played at real-time speed.

**Video S3:** A frame-by-frame video of Voronoi diagram evolution for a segment of the same crystallite motion shown in Video S1, colored according to number of nearest neighbors (5 = blue, 6 = yellow, 7 = red). The video is played at ~ 30x speed and the actual elapsed time is ~ 9 minutes.